\documentclass[%
  reprint,           % PRL 推荐用 reprint 而不是 preprint
  amsmath,amssymb,
  aps, prl,               % 关键：prl
  superscriptaddress,% 作者地址脚注
  longbibliography,  % 可选：显示完整参考文献标题
  floatfix,
  onecolumn
]{revtex4-2}
\usepackage{natbib}
\usepackage{bm}
\usepackage{comment}
\usepackage{color}
\usepackage{bm}
\usepackage{graphicx}
\usepackage{etoolbox}
\usepackage{calc}
\usepackage{amsmath}
\begin{document}

%\preprint{APS/123-QED}

\title{\textbf{Statistics of Thermal Avalanches in Driven Amorphous Systems} 
}% 

\author{Zhiyu Cao}
\affiliation{Center for Theoretical Biological Physics, Rice University, Houston, TX 77005}

\author{Peter G. Wolynes}
\email{pwolynes@rice.edu}
\affiliation{Center for Theoretical Biological Physics, Rice University, Houston, TX 77005}
\affiliation{Department of Chemistry, Rice University, Houston, TX 77005}
\affiliation{Department of Physics, Rice University, Houston, TX 77005}


\date{\today}% It is always \today, today,

%\tableofcontents

%\section*{Significance}
%During mitosis, chromosomes are compacted into cylindrical rod-like structures that ensure their reliable transfer to daughter cells. Using a motorized chromosome model, we underline the roles in mitosis of two specific molecular machines, that can be modeled as grappling motors, condensin I and condensin II. Particular attention is given to the way the operational mechanisms of these motors lead to the breaking of symmetries in mitosis and how their action gives rise to characteristic defects.

\begin{abstract}
Within the framework of the random first-order transition theory of glasses, we discuss the statistics of  ``thermal avalanches", the large scale rearrangements in driven amorphous systems near their instability. Stringy excitations yield non-Poisson waiting‐time statistics. Embedding these statistics in a generalized Master equation captures the non‑Markovian, aging dynamics of avalanche clusters. We apply this framework to analyze nonequilibrium signatures of thermal avalanches—auto-correlation functions and effective temperatures—under both quasi‑static shear and stochastic shaking protocols. We use full counting statistics to derive the complete distribution of both the avalanche magnitudes and avalanche counts, uncovering the intermediate-time behavior.

\end{abstract}

\maketitle
%\tableofcontents
%\newpage
% Use \firstpage to indicate which paragraph and line will start the second page and subsequent formatting. In this example, there are a total of 11 paragraphs on the first page, counting the first level heading as a paragraph. The value {12} represents the number of paragraphs starting on the second page. If a paragraph runs over onto the second page, include a bracket with the paragraph line number starting the second page, followed by the paragraph number in curly brackets, e.g. "\firstpage[4]{11}".

% If your first paragraph (i.e. with the \dropcap) contains a list environment (quote, quotation, theorem, definition, enumerate, itemize...), the line after the list may have some extra indentation. If this is the case, add \parshape=0 to the end of the list environment.
\section{Introduction}
We usually take for granted the solidity of the ground on which we walk. Only occasionally does our footing give way, perhaps when we walk on the side of a steeply inclined hill. Of course, earthquakes also happen rarely. At the molecular level, in contrast, all is in motion all of the time, yet in viscous liquids, local structures do persist for many vibrations \cite{lubchenko2007theory}. In glassy liquids, the persistence times of different regions are widely distributed. We say a liquid has become a glass when most of the rearrangements would take longer than we have been willing to wait to observe them. If we wait a bit longer, however, there is a regime, then where only a few rearrangements typically have time to take place, and these can, in principle, be isolated from each other, counted, and to some extent treated as individuals. On the mesoscale that is, between the molecular and geological length scales, sudden rearrangement events can also often be individuated. In addition, on the mesoscale, systems are easily tuned near to the instability point by external forcing or through their preparation, owing to the weaker effect of thermally activated transitions on such mesoscale objects, which leaves them trapped in quite shallow minima. A completely athermal $T=0$ picture, where driving dominates these rearrangements is usually taken to describe experiments on granular matter \cite{lin2014scaling,tahaei2023scaling,wyart2005effects,wyart2005geometric}. Biological and colloidal systems are, however, in between granular matter and glassy liquids, because both thermally activated motions and driven motions coexist. In this paper, we explore rearrangements on the mesoscale, where both driving and thermal activation play a role, calling these reconfiguration events ``thermal avalanches."

Yielding at the mesoscale at finite temperature presents a unique complexity. Unlike the purely mechanical, athermal ($T=0$) limit typical of granular matter \cite{lin2014scaling,ozawa2018random}, thermal fluctuations are important for colloids and biological systems. The dynamics also cannot be reduced to the study of the typical independent thermal activation events as in a near equilibrium glassy liquid. Rare thermal fluctuations act as triggers for extensive mechanical rearrangements, in these “thermal avalanches” \cite{chauve2000creep,de2025dynamical,tahaei2023scaling,durin2024earthquakelike,ferrero2021creep,korchinski2024microscopic}. Evidence for this hybrid mechanism extends to biological systems, where “cytoquakes” drive cytoskeletal remodeling \cite{zhou2009universal,alencar2016non,liman2020role,floyd2021understanding,wang2012active}, and to soft matter systems \cite{bouchaud2001anomalous,angelini2013dichotomic,gabriel2015compressed,li2019physical,cao2025motorized}. Near the instability to yielding, Wisitsorasak and Wolynes showed within the RFOT theory framework that the strength of glasses is set by stress-catalyzed cooperative rearrangements, where applied stress progressively lowers activation barriers and leads to a spinodal-like onset of barrierless reconfiguration at the intrinsic yield strength \cite{wisitsorasak2012strength}. The theory quantitatively relates yielding to the elastic modulus and to the configurational energy frozen in at glass formation. This framework was later cast into a dynamical, spatially resolved form to account for heterogeneous flow and shear-band formation under deformation \cite{wisitsorasak2017dynamical}.

Computer simulations \cite{kob1997dynamical,donati1998stringlike,donati1999spatial,gebremichael2004particle} and colloidal‐glass experiments \cite{weeks2000three} have shown that cooperative rearrangements near instability regions are “fractal” \cite{reinsberg2002comparative} or “stringy” clusters \cite{donati1998stringlike,gebremichael2004particle,langer2005dynamic}. The random first-order transition (RFOT) theory successfully captures this transition \cite{stevenson2006shapes}: near the laboratory glass transition of molecular liquids, cooperative rearranging regions are compact, but near the higher dynamic crossover temperature $T_c$—at which molecular scale motion ceases to be activated and becomes dominated by mode coupling and facilitation effects \cite{bhattacharyya2008facilitation}—stringy rearrangements prevail. At the dynamic crossover from activated to non-activated motion, local aging events break the particle cages, destabilizing neighbors, thereby triggering cascades of excitations in a cascade of avalanches in adjacent regions \cite{lubchenko2017aging,zhang2021interplay}. Furthermore, these avalanches can branch, forming rearranging particle clusters that resemble percolation clusters. Ultimately, these can overlap leading to flow. While different in apparent starting point from the elasto‐plastic models that describe local plastic relaxation and its long‐range coupling via Eshelby‐type stress kernels \cite{dyre2006colloquium,tahaei2023scaling,picard2004elastic,maloney2006amorphous,lin2014scaling,lemaitre2014structural,bocquet2009kinetic}, the RFOT theory uncovers the same stress‐redistribution mechanisms from the standpoint of free‐energy landscape and entropy‐driven activation. Both approaches are fundamentally consistent and mutually complementary \cite{lubchenko2009shear,rabochiy2012liquid}.

Here, using the RFOT theory framework, we map the stochastic free‐energy landscape of string‐like rearrangements under increasing imposed loads or external driving onto a continuous time random walk (CTRW) by calculating forward and backward transition rates using Metropolis dynamics with an imposed facilitation cutoff, coming from the interaction of activated events \cite{xia2001microscopic,wisitsorasak2014dynamical}. The avalanche sizes are discretized, reflecting the grain-like or particle-scale nature of the rearrangements. The resulting statistics depart from previous continuum theories based on the RFOT theory, which treated stress generating events as local heterogeneous Gaussian random forces that obey fluctuation dissipation relations \cite{bhattacharyya2008facilitation,wisitsorasak2017dynamical}. From the derived rate distributions, we obtain non-exponential waiting‐time distributions for both thermal activation events and externally driven initiation of avalanches, which allows us to formulate a generalized Master equation. We use this CTRW framework to analyze avalanche statistics under diverse driving protocols and explore their nonequilibrium noise characteristics. Beyond finding the average behavior of the growth of avalanches, we introduce a full counting statistics approach \cite{levitov1993charge,levitov1996electron,nazarov2012quantum,bagrets2003full,flindt2008counting,braggio2006full} to understand the statistics of individual events by introducing a counting field is coupled to a CTRW kernel. The resulting kernel deformed equation allows us to find the full counting probability distributions for the intermediate-time and long-time behavior of the avalanche statistics.

%Finally, we extend our framework to multi-step transition processes through a renormalization group analysis, better revealing increasingly cooperative avalanches with heavy-tailed statistics.

%The organization of this paper is as follows: In Sec.~II, we review the RFOT string rearrangements and their mapping to a 1D stochastic process.  Sec.~III introduces the CTRW framework for thermal and driven avalanches, and Sec.~IV applies it to derive the average statistics and effective temperatures under driving protocols, while Sec.~VI examines intermediate‐time aging and full‐counting statistics.  In Sec.~VII, we study the continuum stress‐diffusion model with avalanche‐induced mobility, and Sec.~VIII wraps up with conclusions and outlook.

\section{The Fluctuating Free Energy of RFOT Strings near Yielding}
To understand rearrangement events near instability, RFOT theory balances the drive to explore new configurations arising  from the entropy of accessible configurations and frozen in enthalpy with the interfacial free energy cost mismatch between distinct metastable configurations. The microscopic development of the RFOT theory of glasses based on density functionals suggests this mismatch surface tension has a nearly universal form near $T_g$. The mismatch energy $ \sigma_0\!=\!(3/4)k_{\!B}Tr_0^{-2}\!\ln({1}/{d_L^2\pi e})$\cite{kirkpatrick1989scaling,xia2000fragilities}, where $d_L$ is the Lindemann length. The Lindemann length is nearly universally a tenth of the inter-particle spacing ($d_L \!=\! 0.1r_0$), so $\sigma_0$  only varies weakly from substance to substance. If there is no frozen-in enthalpy, the free-energy cost to rearrange a cluster with size $L$ and surface area $\Sigma$ transiting to a new metastable state is
\begin{equation}
    F(L,\Sigma)\!=\!\Sigma\sigma_0-LTs_c-k_BT\ln\Omega(L,\Sigma)-\!\sum_{particles}\!\tilde{f}, \label{FNS}
\end{equation}
where $s_c$ is the configurational entropy per particle. $\Omega(L,\Sigma)$ counts the number of cluster shapes (near zero for compact droplets, $\propto \!L$ for stringy or percolation‐like structures). The last term accounts for the fact that local energy fluctuates. $\tilde f_i$ is a random driving‐force fluctuation, with r.m.s. width $\delta f\!\approx\! T\delta s_c\!\approx\! T\!\sqrt{\!\Delta C_pk_B}$, where $\Delta C_p$ is the configurational heat capacity of the fluctuating region. 

Near the instability, we can then analyze avalanche free energy statistics using string‐like clusters. Beyond the initiation cost $F_{\text{in}}$, the free energy grows linearly $F(L)\!= \!\phi L \!+ \!F_{\text{in}}$ \cite{stevenson2006shapes}, where $\phi\!=\!T\Delta s_c$ is the slope of the linear string free-energy profile with $\Delta s_c\!= \!s_c^{\text{string}}\!-\!s_c$. The critical string entropy satisfies $Ts_c^{\text{string}}\! =\! \nu_{\text{int}}(z \!- \!2)\! -\! k_{\!B} T \ln (z\! - \!5)\! \approx\! 1.13 k_{B} T$ with $z\! =\! 12$ for the random close-packed lattice ($\nu_{\rm int}$ is the surface tension per neighbor). The critical temperature for the stringy transition locates when \( s_c(T_{\text{string}}) \!=\! s_c^{\text{string}} \). Below \( T_{\text{string}} \) is the uphill case, where pure string reconfiguration is forbidden. Above \( T_{\text{string}} \) is the downhill case, where only \( F_{\text{in}} \) must be overcome. Hence, \( T_{\text{string}} \) marks the crossover from activated to non‑activated dynamics and corresponds to \( T_c \), as confirmed by experiment. Since $s_c(T)=S_\infty(1-T_K/T)$ with $S_\infty=\Delta C_pT_g/T_K$ ($T_g$ and $T_f$ are the glass transition temperature and the Kauzmann temperature, respectively) \cite{richert1998dynamics,stevenson2005thermodynamic}, we have that $\beta\phi\!=\!\Delta s_c\!=\!\Delta C_pT_g(\beta\!-\!\beta_c)$. If frozen-in enthalpy is present, the reconfiguration drive is enhanced by the excess configurational energy stored at preparation. In the RFOT theory, this enters as an additional bulk contribution $\Delta\Phi$ per particle within the fictive-temperature description of nonequilibrium glasses \cite{lubchenko2004theory,lubchenko2007theory}. The effective string slope is therefore renormalized to $\phi_{\rm eff}=T\Delta s_c+\Delta\Phi$, or equivalently $Ts_c\to Ts_c+\Delta\Phi$ in the instability criterion. The condition for barrierless string growth then becomes $Ts_c(T_{\text{string}})+\Delta\Phi = Ts_c^{\text{string}}$, showing that frozen-in enthalpy lowers the apparent instability threshold.

In the presence of an external shear stress, the string free-energy landscape is modified by stress catalysis of rearrangements. Wisitsorasak and Wolynes \cite{wisitsorasak2012strength} showed an applied stress lowers the effective free-energy cost of reconfiguration by releasing elastic strain energy, adding a negative contribution proportional to $\sigma^2 / 2G$ per particle to the driving force. Consequently, the linear slope $\phi$ moves downward under stress and vanishes at a critical stress, signaling a spinodal-like yielding instability where string-like rearrangements become barrierless even below $T_{\text{string}}$. We see that thermal activation and mechanical loading play analogous roles in triggering avalanches.

In their theory of secondary relaxation, Stevenson and Wolynes derived the free-energy barrier statistics for stringy rearrangements below the instability \cite{stevenson2010universal}. Following them, we see that under uncorrelated disorder, Eq. \ref{FNS}, maps to a free-energy random walk. Including fluctuations and taking the continuum limit yields
\begin{equation}
    \frac{dF}{dN}=\phi+\tilde{f}_N. \label{dF/dN}
\end{equation}
where internal noise enables string rearrangements even for $T<T_c$, and makes them especially susceptible to driving near $T_c$.

\section{A formalism of CTRW description for avalanches}
Eq.\eqref{dF/dN} allows us to obtain waiting-time statistics for a CTRW described by a generalized Master equation. This calculation resembles the analysis of Bryngelson and Wolynes of intermediates in a biased random energy model of protein folding dynamics \cite{bryngelson1989intermediates}. Using Metropolis dynamics for the thermal motions, i.e., the transition rate between states $A$ and $B$ is $R_0e^{-\beta(E_B-E_A)}$ when $E_B>E_A$, and $R_0$ for else with $R_0$ the intrinsic transition rate. This model essentially ignores fluctuations in the local part of the barrier heights for individual steps. The cluster size of the growing avalanches performs a 1D random walk with asymmetric rates determined by local barriers. Barriers for backward ($L\!\to\! L\!-\!1$) and forward ($L\!\to\! L\!+\!1$) jumps are 
$$\Delta E^-=-(\phi+\tilde{f}_L),\quad\Delta E^+=\phi+\tilde{f}_{L+1}$$
if $\Delta E^\pm\!\ge\!0$. The distribution for the slow transitions is truncated by an $\alpha$ relaxation cutoff $R_\alpha$ \cite{xia2001microscopic} because such very slow barrier regions will not be able to reorganize on their own but rather will usually be equilibrated by a typical neighboring event facilitating the transition.
\begin{equation}
   {R}^-(\tilde{f}_{L})=\begin{cases}
   R_\alpha& \tilde{f}_{L}<\tilde{f}^-_\alpha\\
R_0e^{\beta(\phi+\tilde{f}_{L})}& \tilde{f}^-_\alpha\le\tilde{f}_{L}<-\phi\\
R_0 & \tilde{f}_{L}\geq-\phi
\end{cases}\label{R-f}
\end{equation}
\begin{equation}
   {R}^+(\tilde{f}_{L+1})=\begin{cases}  
   R_\alpha& \tilde{f}_{L+1}>\tilde{f}^+_\alpha\\
R_0e^{-\beta(\phi+\tilde{f}_{L+1})}& \tilde{f}^+_\alpha\ge\tilde{f}_{L+1}>-\phi\\
R_0 & \tilde{f}_{L+1}\leq-\phi
\end{cases},\label{R+f}
\end{equation}
where $\tilde f^\pm_\alpha=-\phi\pm T\ln(R_0/R_\alpha)$. Using Eqs.\eqref{R-f} and \eqref{R+f} and $g(\tilde f)$ for the Gaussian noise, the thermal avalanche rate distribution becomes
\begin{equation}
     P_{\rm th}^\pm(R)\sim R^{-1}\exp\left[-\left(-\phi\pm T\ln({R_0}/{R})\right)^2/2\delta f^2\right]\label{Pthpm}
\end{equation}
for intermediate rates where $R_\alpha <R<R_0$, but is zero outside that range. The corresponding waiting time distributions then the form:
\begin{equation}
    \psi^\pm_{\rm th}(t)\sim\int_{\eta}^{1}dr \exp\left[-(R_0 t)r-
    \frac{\left(\Delta s_c\pm\ln r\right)^2}{2\Delta C_p}\right]\label{psi1}
\end{equation}
with $\eta=R_\alpha/R_0$ and $r=R/R_0$. These waiting time distributions feed into a generalized Master equation for avalanche dynamics \cite{cox1966theory,feller1991introduction}. By sampling from the waiting time distributions in Eq.\ref{psi1}, we can also directly simulate the dynamical evolution of cluster size $L$. The parameters we selected for our detailed discussions are guided by the experimentally measured branching rates $R_0\!\sim\!0.01-1s^{-1}$ in cytoskeletal systems (one of our target systems) \cite{liman2020role}. 

\section{Protocols in avalanche dynamics}
We explicitly discuss two driving protocols often used in glassy rheology and avalanche studies, athermal quasi‐static loading in which the stress is slowly increased and random “shaking.” In athermal quasi-static simulations and rheometer tests on colloidal glasses and emulsions, a small constant shear rate (or stress) continuously increases loading local regions until yielding triggers avalanches \cite{parisi2017shear,coussot2002avalanche,agoritsas2015relevance,nicolas2018deformation}. For the random “shaking” protocol, acoustic vibrations, or intermittent force pulses induce local stress spikes \cite{nicolas2018deformation,sanz2014avalanches,gibaud2020rheoacoustic}. When these spikes exceed the local yield threshold, they nucleate avalanches which are thus driven by a tunable noise bath.

\subsection{The external shear stress}
\begin{figure}[htbp]
  \centering
  \includegraphics[width=0.7\columnwidth]{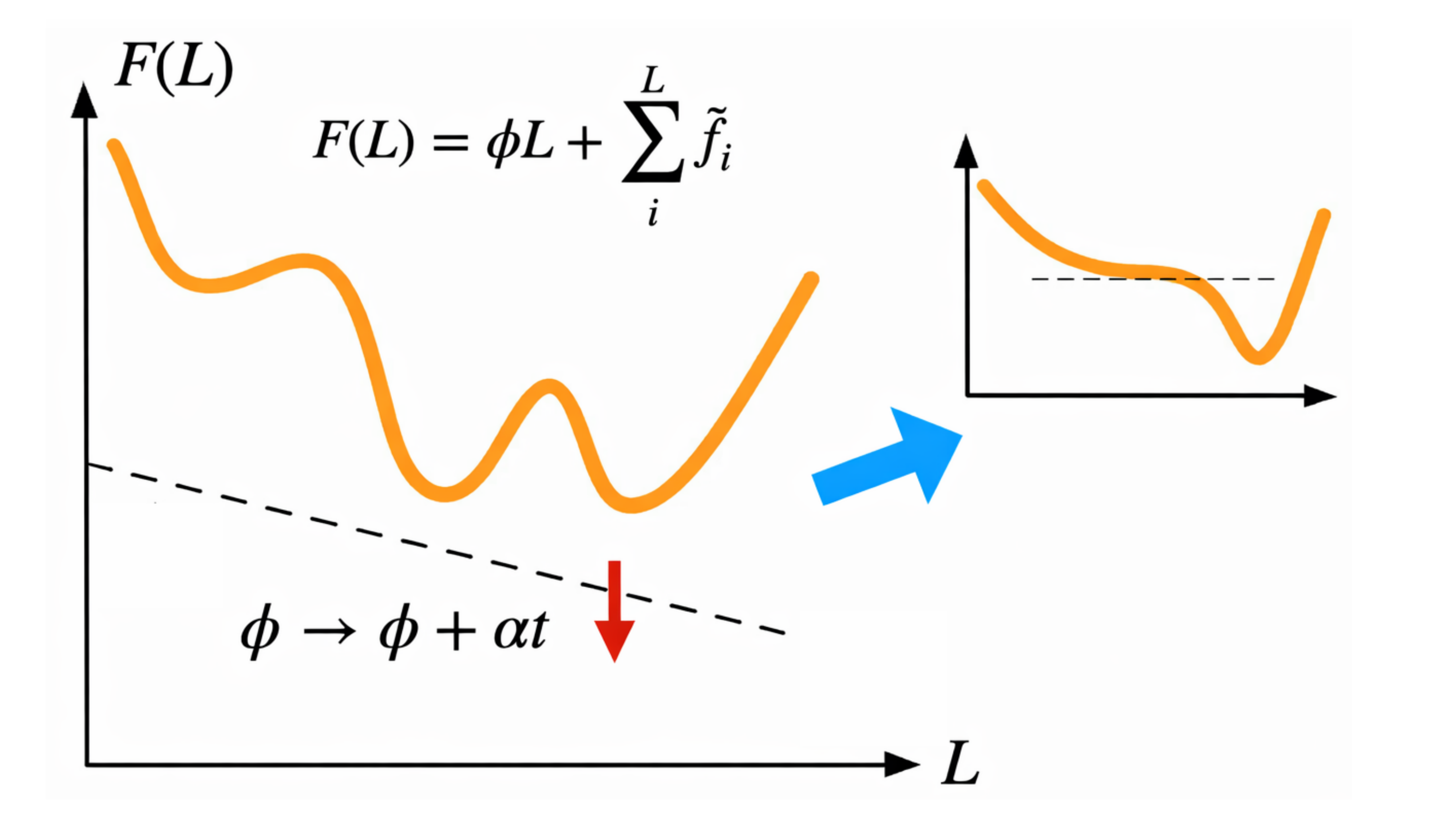}
  \caption{The free energy $F(L)$ of a string of length $L$ consists of a deterministic linear contribution $\phi L$ and quenched fluctuations, leading to a rugged landscape with multiple metastable minima. A slow external ramp, modeled as a linear increase of the driving parameter $\phi \to \phi + \alpha t$, progressively tilts the landscape (dashed line), reducing local barriers. As the tilt increases, metastable states disappear via a spinodal-like instability, triggering irreversible string growth events (avalanches). The inset illustrates the local destabilization of a single metastable minimum under the applied ramp.}
  \label{fig:ramp}
\end{figure}

To capture the progressive stress build-up in our RFOT-CTRW model, we let the driving force evolve in time as $\phi_j=\phi+j\alpha\tau$ to track stress buildup. For simplicity, time is discretized as $\{\tau_0, \tau_1,\dots,\tau_M\}$ with $\tau_j\!=\!j\tau$ ($j\!=\!0,\dots,M$,\! $M\tau\!=\!\mathcal T$). This protocol mimics a quasistatic ramp of external stress, under which metastable configurations are destabilized sequentially rather than simultaneously, allowing avalanches to be triggered by the disappearance of individual local minima, see Fig. \ref{fig:ramp}.

Using this discretization, minima disappear as the driving force is increased serially. A cluster at $L$ is a local minimum if $F(L)\!<\!F(L-1)$ and $F(L)\!<\!F(L+1)$, i.e.,
$$\phi_j\!+\!\tilde{f}_L\!<\!0,\quad\phi_j\!+\!\tilde{f}_{L+1}\!>\!0.$$
For initial stability ($\tilde f_{L+1}>-\phi$), the second inequality flips at
\begin{equation}
    t_c\!=\!-\frac{(\phi\!+\!\tilde{f}_{L+1})}{\alpha}.\label{tc}
\end{equation}
Physically, this time corresponds to a spinodal-like loss of local stability, where the barrier vanishes and irreversible string growth is initiated.

Treating $\tilde f$ as Gaussian noise, these gives the avalanche initiation waiting time distribution
\begin{equation}
    \psi_{\rm av}^+(t)\sim\exp\left[-\frac{(\phi+\alpha t)^2}{2\delta f^2}\right].
\end{equation}
This $\psi^+_{\rm av}(t)$ then truncates the forward waiting time distribution at high barriers. The combined forward waiting time distribution is automatically normalized,
$$\psi^+(t) = \psi^+_{\rm th}(t)S_{\rm av}^+(t)+\psi_{\rm av}^+(t)S_{\rm th}^+(t),$$
where $S_{\rm av}^+(t)\!=\!\int_t^\infty\!\psi_{\rm av}^+(t)dt$ and $S_{\rm th}^+(t)\!=\!\int_t^\infty\!\psi_{th}^+(t)dt$ are the survival probabilities. Note that initiation events enter only the “+” branch, so that the backward waiting time distribution $\psi^-(t)\!=\!\psi_{\rm th}^-(t)$. The overall waiting time fluctuations are 
$$\psi_{\rm eff}(t)=\psi^+(t)S^+(t)+\psi^-(t)S^+(t)$$
with $S^+(t)\!=\!\int_t^\infty\psi^+(t)$ and $S^-(t)\!=\!\int_t^\infty\psi^-(t)$. Non‐exponential kernels thus produce coupled relaxation channels \cite{nandi2019continuous,seki2023fluctuation}. The generalized Master equation corresponding to the CTRW with thermal ($\pm$) and avalanche ($+$) channels is
\begin{equation}
    \begin{aligned}
        \frac{\partial P_L(t)}{\partial t}
&=\!\int_0^t\!d\tau[
K^+\!(t-\tau)P_{L-1}(\tau)\!+\!K^-(t-\tau)P_{L+1}(\tau)\\
&+\!K_{\rm av}(t-\tau)P_{L-1}(\tau)
]\!-\!\int_0^t\!d\tau K_{\rm tot}(t-\tau)P_L(\tau).
    \end{aligned} \label{GME_stress}
\end{equation}
Here, $K^\pm$ and $K_{\rm av}$ are the memory kernels of the thermal movements and initiated avalanches, and $K_{\rm tot}\!=\!K^+\!+\!K^-\!+\!K_{\rm av}$. The memory kernels can be derived from the waiting time distributions through the Montroll-Weiss-like formula.

\subsection{The random shaking}
We also analyze a random shaking protocol, in which alternating thermal‐freeze cycles induce aging, rejuvenation, and memory effects \cite{peter2009spatiotemporal,berthier2011theoretical,chacko2024dynamical}.  This protocol is equivalent to there being a walker that occupies two parallel lattices coupled to baths at different temperatures. On each lattice, forward and backward hops follow the waiting time distributions $\psi_1^\pm(t)$ or $\psi_2^\pm(t)$, and inter‐lattice transfers occur via $\psi^\delta(t)$ or $\psi^\gamma(t)$. This setup maps straightforwardly to a parallel‐chain CTRW governed by coupled generalized Master equations \cite{flomenbom2005closed,stukalin2006transport,teimouri2013all}
\begin{equation}
\begin{aligned}
    \frac{\partial P_{j,1}(t)}{\partial t}\!=&- \!\int_{0}^t\!dt' \!\big[K^+_1(t')+K^-_1(t')+K^\gamma(t')\big]\!P_{j,1}(t-t')\\
    &+\int_{0}^tdt' \big[K^+_1(t') P_{j-1,1}(t - t')\\
    &+K^-_1(t') P_{j+1,1}(t-t')+K^\delta(t')P_{j,2}(t - t')\big] , \label{GME1}
\end{aligned}
\end{equation}
and
\begin{equation}
\begin{aligned}
    \frac{\partial P_{j,2}(t)}{\partial t}\!=&- \!\int_{0}^t\!dt' \!\big[K^+_2(t')+K^-_2(t')+K^\delta(t')\big]\!P_{j,1}(t-t')\\
    &+\int_{0}^tdt' \big[K^+_2(t') P_{j-1,1}(t - t')\\
    &+K^-_2(t') P_{j+1,1}(t-t')+K^\gamma(t')P_{j,2}(t - t')\big] , \label{GME2}
\end{aligned}
\end{equation}
Here, \( P_{j,l}(t) \) is the probability of finding the system in a lattice $l=1,2$  with $j$ particles at time \( t \). $K_l^\pm(t)$ and $K^{\delta,\gamma}(t)$ are the memory kernels, which can be derived from the waiting time distributions:
\begin{equation}
\tilde{K}_\alpha^\pm(s)=\frac{s\tilde{\psi}_\alpha^\pm(s)}{1-\tilde{\psi}_\alpha(s)};\quad \tilde{K}^\gamma(s)=\frac{s\tilde{\psi}^\gamma(s)}{1-\tilde{\psi}_1(s)};\quad\tilde{K}^\delta(s)=\frac{s\tilde{\psi}^\delta(s)}{1-\tilde{\psi}_2(s)}
\end{equation}
where $\alpha\in\{1,2\}$, $\tilde{\psi}_1(s)=\tilde{\psi}_1^+(s)+\tilde{\psi}_1^-(s)+\tilde{\psi}^\gamma(s)$ and $\tilde{\psi}_2(s)=\tilde{\psi}_2^+(s)+\tilde{\psi}_2^-(s)+\tilde{\psi}^\delta(s)$. 

\section{The static average statistics and the effective temperature}
To characterize the long-time statistics of avalanche activity, we focus on the fluctuations of the cumulative growth of a single initiated avalanche $L(t)$ over a time interval $t$. While the average growth rate captures the typical dynamics, rare fluctuations away from the mean play a crucial role in nonequilibrium systems and cannot be described by standard central-limit arguments alone. The large deviation principle provides a systematic framework to quantify the probability of such atypical trajectories by associating them with an effective free-energy–like rate function. Within this framework, the probability distribution of the time-averaged activity $L(t)/t$ obeys an exponential scaling in time, analogous to Boltzmann weights in equilibrium statistical mechanics. This analogy allows one to define a generating function and its Legendre transform, from which both the typical behavior and fluctuation-induced corrections can be extracted in a unified manner.

According to the large deviation principle analysis \cite{touchette2009large}, then, we have that $P({L(t)}/{t} \approx v)\! \sim\! e^{-t I(v)}$. Here, $I(v)$ is the rate function, which can be calculated from the scaled cumulant generating function $\mu(\lambda)\sim t^{-1}\ln\langle e^{\lambda L(t)}\rangle$ as
\begin{equation}
    I(v) = \sup_{\lambda} \left[\lambda v - \mu(\lambda)\right].
\end{equation}
The long time limit will yield a Gaussian approximation $I(v)\! \approx\! {(v - v_{\text{avg}})^2}/{2 D}$ near its minimum, where the average velocity and the diffusivity are
\begin{equation}
    v_{\text{avg}} = \mu'(0),\quad D=\mu''(0)/2. \label{vD}
\end{equation}
Using Eq. \ref{vD}, we find that (see \textcolor{blue}{SI} for details)
\begin{equation}
    v_{\text{avg}} = \frac{\Delta p}{\langle \tau \rangle},\quad D = \frac{1 - \Delta p^2}{2\langle \tau \rangle} +\frac{\Delta p^2 \sigma_\tau^2}{2\langle \tau \rangle^3}, \label{D}
\end{equation}
where $p_\pm$ the probabilities for forward and backward transitions with $\Delta p=p_+-p_-$ the drift. $\langle\tau\rangle$ and $\sigma_\tau^2$ are the mean and the variance of the overall waiting time fluctuations, which contains the contributions from all the channels. Eq.\ref{D} reveals two diffusion sources: directional randomness (first term) and waiting‐time variability (second term), the latter dominating for heavy‐tailed kernels \cite{akimoto2018ergodicity}. 

The avalanche dynamics is nonequilibrium due to the external driving. An effective temperature quantifying these nonequilibrium characteristics of the steady states can be defined via an Einstein-like relation \cite{he2008random,hou2018biased}:
\begin{equation}
    T_{\rm eff} \;\equiv\; \frac{D}{\mu},\label{Teff}
\end{equation}
where $\mu$ is the mobility (how the velocity varies with driving force). The analysis shows in the linear response region with small driving forces $\phi$, $ T_{\rm eff}/T\!\to\!1$, but stronger driving yields $T_{\rm eff}\neq T$ \cite{seki2023transient}. 

For the shear stress protocol, $\psi^+$ include both the thermal motions and the initiation of avalanches induced by external shear. As discussed earlier, the external ramping will effectively truncate the long-waiting-time moves in the forward transition. This will result in a correction of the drift. In the limit of relatively slow tuning of the stress level $\phi\!\gg\!|\alpha|\tau\!>\!0$ without loss of generality, the drift at the long time limit $t\gg \phi/|\alpha|$ can be calculated as (see \textcolor{blue}{SI} for details)
$$\Delta p\simeq\Delta p_0
          +\bigl(1-\Delta p_0\bigr)
           \exp\!\Bigl[-\frac{\phi R_0}{|\alpha|} \bigl(1+e^{-\beta\phi}\bigr)\Bigr]$$
with $\Delta p_0\!=\!\tanh(-\frac{\beta\phi}{2})\!<\!0$ the drift without external shear stress. The second term represents the reduction of the net backward displacement due to the truncation of long–waiting–time forward jumps from the avalanche channel. Using Eq.\ref{Teff}, we have that  
\begin{equation}
    T_{\rm eff}\simeq T\Bigl(1+\frac{2e^{-\phi R_0/|\alpha|}}{\Delta s_c}\Big). \label{Te/T}
\end{equation}
This expression holds only when the external stress tuning rate $|\alpha|$ is small. When the external stress is absent, $\phi/|\alpha|\to\infty$, therefore $T_{\rm eff}$ reduces to the thermal temperature $T$. When a small shear stress is introduced, the initiation of avalanches contributes a correction leading to a higher effective temperature, showing there is an effective heating induced by avalanches, which we point out is not directly due to the released potential energy which we assume to be dissipated.

The same calculations for the velocity and broadening diffusion at long times can be directly generalized to the random shaking protocol with parallel waiting time distributions for different baths. We model the system as alternating between baths at $T_h$ and $T_l$ with switching rates $\gamma_{l\to h}$ and $\gamma_{h\to l}$, and the stationary weights can be defined as \(\rho_{h}\!=\!\gamma_{l\to h}/\Gamma\) and \(\rho_{l}\!=\!\gamma_{h\to l}/\Gamma\) with $\Gamma\!=\!\gamma_{h\to l}\!+\!\gamma_{l\to h}$ and \(\rho_{l}\!=\!1\!-\!\rho_{h}\). The temperature difference is 
$$\Delta\beta=\beta_l-\beta_h>0.$$
The difference of driving forces between two baths is $\Delta s_{c,h}-\Delta s_{c,l}=-\Delta C_pT_g\Delta\beta$.  This ``blinking” bath delivers a continuous heat current \(T_{h}\!\to\! T_{l}\), breaking equilibrium much like Brownian ratchets. In the long time limit for \(t\gg\Gamma^{-1}\), one finds (see \textcolor{blue}{SI} for details):
\begin{equation}
    T_{\rm eff}\approx T_{\rm mix}
        \Bigl[1+\frac{\rho_h\rho_l(\Delta C_pT_g)^2}{8}\,\Delta \beta^2
      \frac{R_0}{\Gamma}
      +\mathcal O(\Delta \beta^4)\Bigr], \label{Te/Tm}
\end{equation}
where \(T_{\rm mix}\!\equiv\!(\rho_{h}\beta_{h}\!+\!\rho_{l}\beta_{l})^{-1}\) is the equilibrium mixed (``blended”) temperature of the two baths.  The positive $\Delta \beta^2$ correction vanishes only if the two baths have the same temperature $T_h=T_l$ or the switching is infinitely rapid at the adiabatic limit $\Gamma/R_0\!\to\!\infty$. Therefore, the classical Einstein equality breaks down in a genuine nonequilibrium steady state, yielding an effective temperature $T_{\rm eff}\!>\!T_{\rm mix}$. Each $T_l\!\to\! T_h$ switch injects energy, while $T_h\!\to\! T_l$ extracts energy dissipation related to the excess entropy production \cite{hatano2001steady} (can be measured via the magnitude of Einstein relation violation \cite{harada2005equality}), yielding a net dissipation proportional to $(T_h\!-\!T_l)^2$. This $(\nabla T)^2$ scaling parallels entropy‐production analyzes for a particle subject to a continuous temperature gradient \cite{celani2012anomalous}.

In nonequilibrium cytoskeletal networks driven by steady shear or by rapid alternation between hot- and cold-ATP baths, molecular motors inject bursts of work that inflate the tracer’s displacement faster than they renormalize its low-frequency mobility. The resulting breakdown of the Einstein relation inevitably leads to a fictive temperature $T_{\rm eff}$ higher than the equilibrium one \cite{shen2004stability,wang2011communication}. 
The $T_{\rm eff}$ expressions for both shear stress protocol and random shaking protocol collapse to the equilibrium $T$ or $T_{\rm mix}$ at vanishing drive yet rise hyperbolically (shear) or quadratically (shaking) with the control parameter, reflecting how stress biases or temperature jumps shorten the long waiting times that normally damp forward motion.

\begin{figure}[htbp]
  \centering
  \includegraphics[width=1\columnwidth]{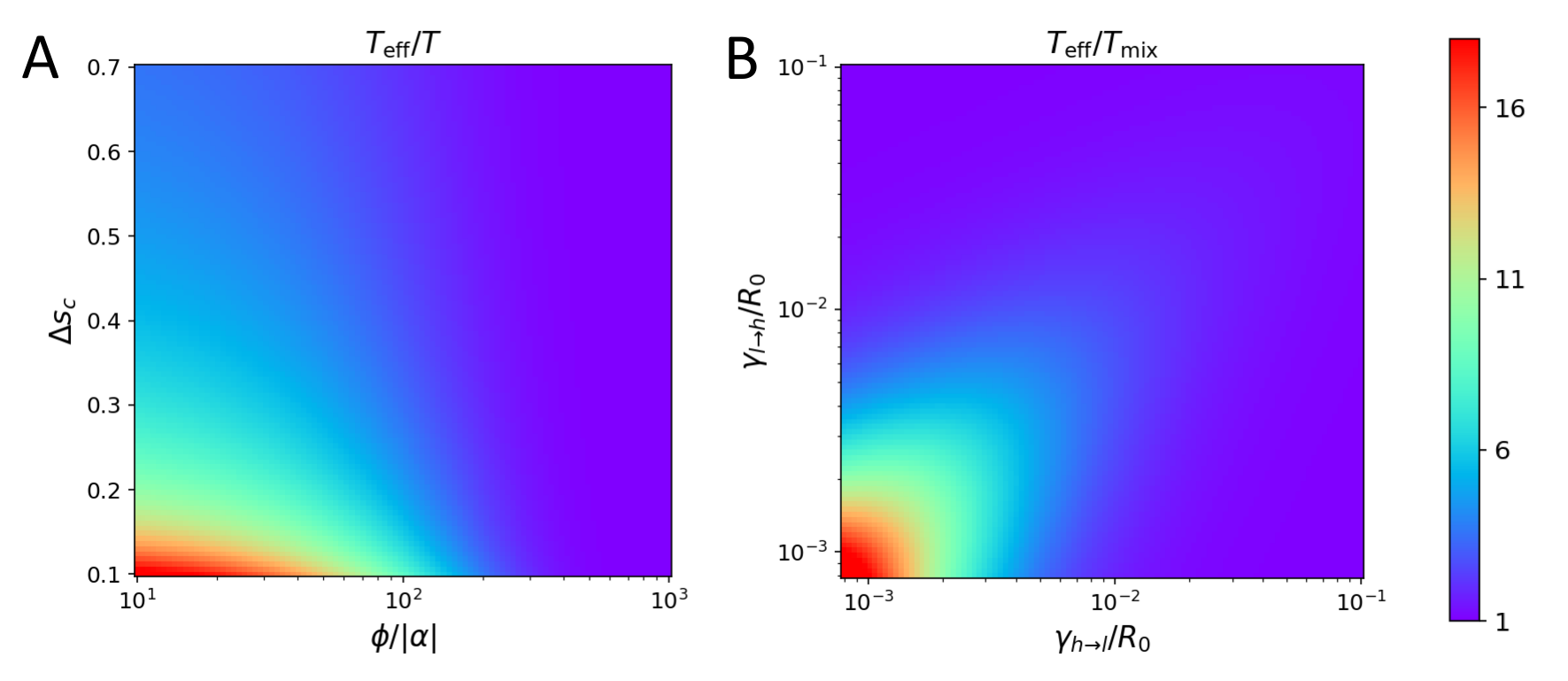}
  \caption{Effective temperature $T_{\rm eff}$ under different protocols. (A) The external shear stress. $T_{\rm eff}/T$ is plotted versus $\Delta s_c$ and $\phi/|\alpha|$. (B) The random shaking protocol. $T_{\rm eff}/T_{\rm mix}$ is plotted versus the switching rates $\gamma_{l\to h}$ and $\gamma_{h\to l}$. Parameters are  $\Delta C_p=1k_b$, $\eta=0$ and $R_0=0.01s^{-1}$. $T_g=250K$ is the glass transition temperature for cross-linked networks \cite{sharaf1980effects}, $T_l=T_g$ and $T_h=2T_g$. }
  \label{fig:Teff}
\end{figure}

In Fig. \ref{fig:Teff}, we plot the effective temperature $T_{\rm eff}/T$ (Eq.\ref{Te/T}) and $T_{\rm eff}/T_{\rm mix}$ (Eq.\ref{Te/Tm}) for both the shear stress tuning protocol and the random shaking protocol, respectively. The chosen parameters are comparable to those relevant to cytoskeletal experiments \cite{zhou2009universal,liman2020role,fritzsche2013analysis}. The calculated effective temperatures can reach more than ten times that of the equilibrium counterpart, as predicted by previous studies of the active cytoskeletal systems \cite{loi2011effective,loi2011non}. This elevation is not merely theoretical: reconstituted actomyosin gels driven by myosin motors display $T_{\rm eff}$ values one to four orders of magnitude above bath temperature when measured with optical tweezers or magnetic rheometers, collapsing back to $T$ after ATP depletion \cite{mizuno2007nonequilibrium}. Likewise, a quadratic increase of  $T_{\rm eff}$ with external driving $f$ in the experimental active colloids \cite{loi2011effective,ginot2015nonequilibrium,palacci2010sedimentation,demery2019driven}, mirroring the $\Delta \beta^{2}$ correction we find for the random shaking protocol.

\section{The intermediate time behavior: auto-correlation function and counting statistics}
\subsection{The auto-correlation function}
Avalanching systems age. We can calculate the auto-correlation function of the growth to investigate the system’s relaxation behavior at intermediate timescales \cite{verberk2003photon,margolin2004aging,witkoskie2006aging}
\begin{equation}
C(t_w,t)=\frac{\langle L(t_w)L(t_w+t)\rangle-\langle L(t_w)\rangle\langle L(t_w+t)\rangle}{\langle [L(t_w+t)-L(t_w)]^2\rangle},\label{Cov}  
\end{equation}
where $t_w$ is the aging time and $t$ is the lag time. 

\begin{figure}[htbp]
  \centering
  \includegraphics[width=1\columnwidth]{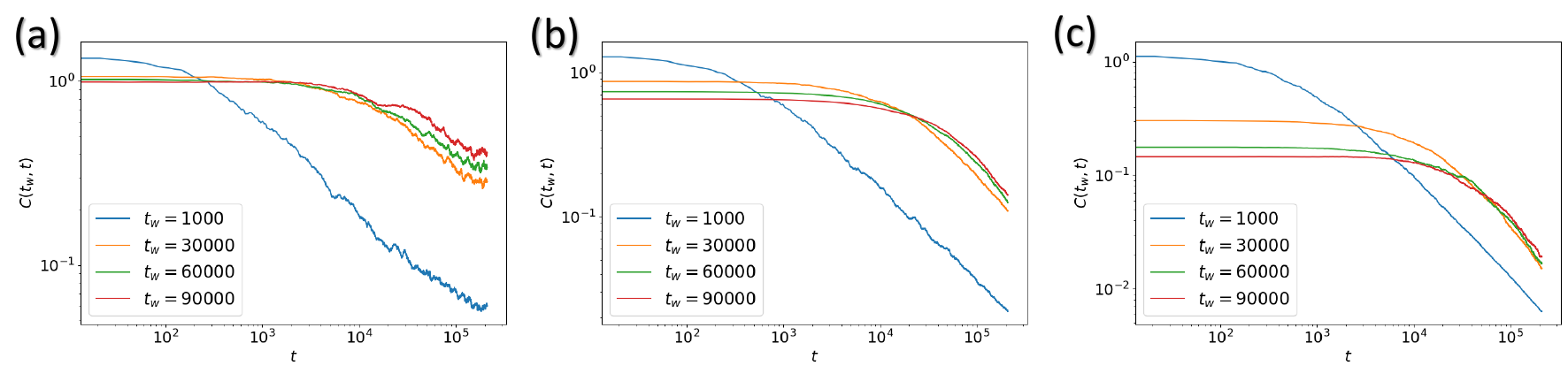}
  \caption{The simulation results of the aging correlation function $C(t_w,t)$ as a function of the lag time $t$ for different aging time $t_w$. Here, $s_c^{\text{string}}=1.13k_B$. (a) $s_c=1.03k_B$; (b) $s_c=1.13k_B$; (a) $s_c=1.23k_B$. Other parameters are $\Delta C_p=1k_b$, $\eta=0$ and $R_0=0.01s^{-1}$.}
  \label{fig:corr}
\end{figure}

Eq.\ref{Cov} can be derived via the double Laplace inversion $t_w\to s$ and $t\to u$. For histories satisfying the condition $L_{tot}(t_w)=n$, we have that $\langle e^{i\lambda L(t_w)}|L_{tot}(t_w)=n\rangle=[G(\lambda)]^n$ with the generating function $G(k)=p_+e^{i\lambda}+p_-e^{-i\lambda}$. Thus, for the unconditional generating function
$$\Phi_L(t_w,\lambda)=\langle e^{i\lambda L(t_w)}\rangle=\sum_{n=0}^\infty P(n,t_w)[G(k)]^n$$
By using the Montroll-Weiss formula, we find the Laplace transform of the correlation
\begin{equation}
\tilde{\Phi}_L(s,\lambda)=\frac{1-\psi_{\mathrm{eff}}(s)}{s[1-G(\lambda)\psi_{\mathrm{eff}}(s)]}.
\end{equation}
From this, the moments can be obtained from
\begin{equation}
    \langle L^{n}(t_w)\rangle=\frac{\partial \tilde{\Phi}_L(s,\lambda)}{\partial (i\lambda)^n}\bigg |_{\lambda=0}.
\end{equation}
The mean displacement is
\begin{equation}
\langle L(t_w)\rangle=\mu\langle L_{tot}(t_w)\rangle, \quad \mathcal{L}\{\langle L_{tot}(t_w)\rangle\}(s)=\frac{\psi_{\mathrm{eff}}(s)}{s[1-\psi_{\mathrm{eff}}(s)]}.
\end{equation}
The second moment is
\begin{equation}
\langle L(t_w)^2\rangle=\Delta p^2\langle L_{tot}(t_w)^2\rangle+(1-\Delta p^2)\langle L_{tot}(t_w)\rangle,
\end{equation}
with
\begin{equation}
\mathcal{L}\{\langle L_{tot}(t_w)^2\rangle\}(s)=\frac{\psi_{\mathrm{eff}}(s)[1+\psi_{\mathrm{eff}}(s)]}{s[1-\psi_{\mathrm{eff}}(s)]^2}.
\end{equation}
We can also derive the two-time joint moment. Using renewal independence, the process can be decomposed into $L(t_w+t)=L(t_w)+\Delta L$, where $\Delta L=\sum_{i=L_{tot}+1}^{L_{tot}(t_w+t)}\xi_i$. Since the renewal process is memoryless, we have
\begin{equation}
\langle L(t_w)L(t_w+t)\rangle=\langle L(t_w)^2\rangle+\langle L(t_w)\rangle\langle\Delta L\rangle,
\end{equation}
with $\langle\Delta L\rangle=\Delta p\langle L_{tot}(t_w+t)-L_{tot}(t_w)\rangle$. Thus,
\begin{equation}
\mathrm{Cov}[L(t_w),L(t_w+t)]=\Delta p^2[\langle L_{tot}(t_w+t)^2\rangle-\langle L_{tot}(t_w)\rangle\langle L_{tot}(t_w+t)\rangle]+(1-\Delta p^2)\langle L_{tot}(t_w)\rangle
\end{equation}
with 
\begin{equation}
    \mathcal{L}_{t_w\to s, t\to u}\left[\langle L_{tot}(t_w) L_{tot}(t_w+t)\rangle\right]=\frac{1}{s}\left[\frac{\psi_{\mathrm{eff}}(s)}{1-\psi_{\mathrm{eff}}(s)}-\frac{\psi_{\mathrm{eff}}(s+u)}{1-\psi_{\mathrm{eff}}(s+u)}\right]\frac{1}{1-\psi_{\mathrm{eff}}(u)},
\end{equation}
and 
\begin{equation}
    \mathcal{L}_{t}\left[L_{tot}(t_w+t)-\langle L_{tot}(t_w)\rangle\right]=\frac{\psi_{\mathrm{eff}}(u)}{u[1-\psi_{\mathrm{eff}}(u)]}.
\end{equation}
Similarly, we can follow the above procedure to get the variance of the increment $\Delta L$:
\begin{equation}
\mathrm{Var}[\Delta L]=\Delta p^2\mathrm{Var}[L_{tot}(t_w+t)-L_{tot}(t_w)]+(1-\Delta p^2)\langle L_{tot}(t_w+t)-L_{tot}(t_w)\rangle.
\end{equation}
Finally, we can obtain the correlation function in the Laplace space:
\begin{equation}
C(t_w,t)=\frac{\mathcal{L}^{-1}[\tilde{C}_{cov}(s,u)]}{\mathcal{L}^{-1}[\tilde{V}(u)]}, \label{CovL}
\end{equation}
where the numerator is the double Laplace transformation of the covariance,
\begin{widetext}\begin{equation}
    \tilde{C}_{\rm cov}(s,u)\!=\!\Delta p^2\!\left\{\frac{1}{s}\left[\frac{\psi_{\mathrm{eff}}(s)}{1-\psi_{\mathrm{eff}}(s)}-\frac{\psi_{\mathrm{eff}}(s+u)}{1-\psi_{\mathrm{eff}}(s+u)}\right]\frac{1}{1-\psi_{\mathrm{eff}}(u)}-\frac{\psi_{\mathrm{eff}}(s)}{s}\frac{\psi_{\mathrm{eff}}(u)}{u[1-\psi_{\mathrm{eff}}(u)]}\right\}\!+\!(1-\Delta p^2)\frac{\psi_{\mathrm{eff}}(s)}{s}
\end{equation} \end{widetext}
and the denominator is the Laplace transformation of the increments of variance 
\begin{equation}
\begin{aligned}
  \tilde{V}(u)
  &=\!\Delta p^2\!\left\{\!
    \frac{\psi_{\mathrm{eff}}(u)\bigl(1+\psi_{\mathrm{eff}}(u)\bigr)}
         {u\bigl[1-\psi_{\mathrm{eff}}(u)\bigr]^2}
    \!-\!\left[\frac{\psi_{\mathrm{eff}}(u)}{u\bigl[1-\psi_{\mathrm{eff}}(u)\bigr]}\right]^2
  \!\right\}\\
  &+\!(1-\Delta p^2)\frac{\psi_{\mathrm{eff}}(u)}{u\bigl[1-\psi_{\mathrm{eff}}(u)\bigr]}.
  \end{aligned}
\end{equation}
In Fig.\ref{fig:corr}, this auto-correlation function is evaluated numerically. We see there is a plateau at the short lag times $t$, reflecting the “freezing” effect during waiting periods. This plateau exists even in the absence of external driving forces, attributable to local barriers, so that the system remains highly correlated with its initial state over short times. With larger $t$, the initial plateau gives way to a power-law decay, indicating a broad spectrum of relaxation times due to random traps. In addition, for large $t_w$, more trajectories stay trapped, and the plateau becomes extended since aging raises the likelihood of long waits.

\subsection{The counting statistics}
We have studied how an avalanche progresses. This progress occurs through a number of individual events. The full distribution of avalanche counts at intermediate timescales—as well as all of their cumulants—can be obtained using full counting statistics \cite{levitov1993charge,levitov1996electron,nazarov2012quantum,bagrets2003full,flindt2008counting,braggio2006full}. In Laplace space, Eq.\ref{GME_stress} can be written as
$$\begin{aligned}
        s\widetilde P_L(s)-P_L(0)
=\widetilde K_1(s)\,\widetilde P_{L-1}(s)
+\widetilde K_2(s)\,\widetilde P_{L+1}(s)
\\+\widetilde K_3(s)\,\widetilde P_{L-1}(s)
-\widetilde K_{\rm tot}(s)\,\widetilde P_L(s)\,.
\end{aligned}$$
To count the different kinds of motions, we then introduce a conjugate counting field $\chi$ for each avalanche induced motion. This quantity acts by deforming the kernel $ \widetilde K_3(s) $ to a complex kernel $ e^{i\chi}\widetilde K_3(s)$. The resulting generating function
$$
G_{L}(\chi,t)=\sum_{l_a} e^{i\chi l_a}P_L(l_a,t)
$$
is the Laplace transform of the count of driven avalanche moves. To obtain the generating function, we write down the complex-deformed generalized Master equation in the Laplace domain as
\begin{equation}
\begin{aligned}
   s\widetilde G_L(\chi,s)\!-\!G_L(\chi,0)
\!=\!\bigl[\widetilde K_1(s)+e^{i\chi}\widetilde K_3(s)\bigr]
\widetilde G_{L-1}(\chi,s)
\\+\widetilde K_2(s)\,\widetilde G_{L+1}(\chi,s)
-\widetilde K_{\rm tot}(s)\widetilde G_{L}(\chi,s).
\end{aligned} \label{CDGME_L}
\end{equation}
In Eq.\ref{CDGME_L}, we can study the progress dynamics by defining the joint generating function $\widetilde G(k,\chi,s)\!=\!\sum_{\!L} \!e^{ikL}\widetilde G_{\!L}(\chi,s)$, which is 
\begin{equation}
\widetilde G(k,\chi,s)
=\bigl[s-\lambda(k,\chi,s)\bigr]^{-1}G(k,t=0). \label{CDGME_S}
\end{equation}
Here, $\lambda(k,\chi,s)\!=\!
\bigl[\widetilde K_1(s)\!+\!e^{i\chi}\widetilde K_3(s)\bigr]e_{+ k}
         \! +\!\widetilde K_2(s)e_{- k}$  with $e_{\pm k}\!=\!e^{\pm ik}\!-\!1$. 
         
Eq.\ref{CDGME_S} provides the full counting statistics of avalanches. We next focus on the conditional propagator $P(L,t|l_a)\!=\!\Pr\bigl\{L(t)\!=\!L\bigl|l_a(t)\!=\!l_a\bigr\}$, which describes how far the system has gone when we know that precisely  $l_a$ avalanches have occurred. Using Eq.\ref{CDGME_S}, this quantity can be obtained through contour integration:
\begin{equation}
P(L,t|l_a)\;=\;
              \frac{\displaystyle\frac{1}{2\pi i}\oint_{|z|=\epsilon}\!\frac{dz}{z^{l_a+1}}\;G_L(z,t)}
                   {\displaystyle\frac{1}{2\pi i}\oint_{|z|=\epsilon}\!\frac{dz}{z^{l_a+1}}\;G_{\rm tot}(z,t)} \label{PN|n}          
\end{equation}
with $G_{\rm tot}(z,t)=\sum_{\!L}\!G_L(z,t)$ and $z=e^{i\chi}$. To all orders, the raw conditional moments can be found using $\langle L^{m}\rangle_{\,\!|l_a}
      ={\partial_{ik}^{\,m}\,\partial_{z}^{\,l_a}\,G(k,z,t)|_{k=0,z=0}}/
                 {\partial_{z}^{\,l_a}\,G(k,z,t)|_{k=0,z=0}}$. Replacing $G$ with $\ln G$ in these derivations yields the cumulants. 

\begin{figure}[htbp]
  \centering
  \includegraphics[width=0.8\columnwidth]{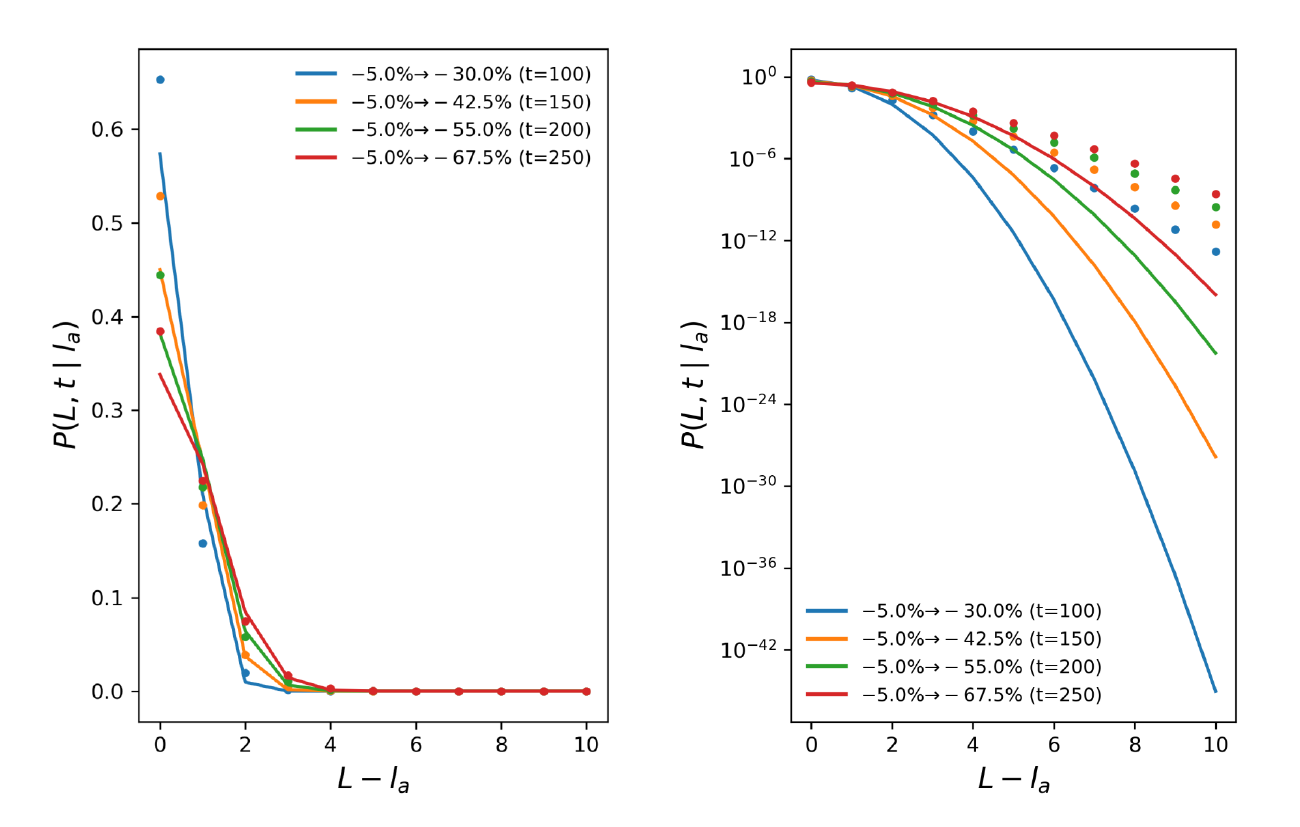}
  \caption{The conditional distribution of avalanches $P(L,t|l_a)$ when the number of avalanche initiations is $l_a=1$: Left panel: linear–linear plot; Right panel: semi-logarithmic plot with a linear horizontal axis and a logarithmic vertical axis. Dots: the full counting statistics [Eq.\ref{Bessel}]; lines: the Gaussian limit [Eq.\ref{Gauss}]. The system is prepared in the downhill flow situation with a frozen configurational state having $s_c^{\rm frozen}=1.1865k_B$ with the critical string entropy $s_c^{\rm string}=1.13k_B$. At the initial moment, the system was $-5\%$ away from instability. The free energy continued to be more and more downhill steadily under the same ramping rate. Over a longer observation period ($t\in\{100,150,200,250\}$), the system moved further away from the instability point (from $-30\%$ to $-67.5\%$).
Other parameters are $\Delta C_p=1k_b$, $\eta=0$ and $R_0=0.01s^{-1}$.}
  \label{fig:PLla}
\end{figure}

We first study the counting statistics for the external shear stress ramp case. The conditional generating function of the conditional propagator $ \widehat G_{l_a}(k,s) \equiv \sum_{L} e^{ikL}\!\int_0^\infty e^{-st} P(L,t| l_a)dt$ becomes
\begin{equation}
\begin{aligned}
     &\widehat G_{l_a}(k,s)= e^{ik l_a} s^{l_a+1}\\
&\times\sum_{r=0}^{\infty} \binom{l_a+r}{r}
\frac{[(1-E(s))(R_0e^{-\beta\phi+ik}+R_0e^{-ik})]^r}{[s+R_0(1+e^{-\beta\phi})]^{l_a+r+1}},
\end{aligned}
\end{equation}
where $E(s)\equiv \exp[-(s+R_0(1+e^{-\beta\phi}))t_*]$ with $t_*={\phi}/{|\alpha|}\gg1$ the timescale of the avalanche initiation. We focus on the intermediate-time regime $R_0^{-1}\ll t\ll e^{R_0 t_*}R_0^{-1}$, where the Laplace transform can be approximated as
\begin{equation}
\begin{aligned}
    P(L,t| l_a)&\approx 
e^{-R_0(1+e^{-\beta\phi})(1-Q)\Delta t}
e^{-\frac{1}{2}\beta\phi(L-l_a)}\\
&\times
I_{|L-l_a|}\!\Bigl(2\Delta t R_0(1-Q)e^{-\frac{1}{2}\beta\phi}\Bigr),\label{Bessel}
\end{aligned}
\end{equation}
where $\Delta t = t - l_a t_*$, $Q=E(0)=\exp[-R_0(1+e^{-\beta \phi})t_*]$ and $I_\nu$ is the modified Bessel function of the first kind. In the long-time limit, the argument of the Bessel function becomes large, while its order $|L-l_a|$ remains of order $\sqrt{\Delta t}$. In this regime, the uniform asymptotic expansion of $I_\nu(x)$ yields a Gaussian form, reflecting the emergence of effective diffusive dynamics from the underlying biased random walk. Due to this Bessel asymptotic, $P(L,t|l_a)$ converges to a Gaussian decay in the long time limit, which can be described by the average velocity and diffusivity (c.f. Eq. \ref{D})
\begin{equation}
    P(L,t| l_a)\to\frac{1}{\sqrt{4\pi D\Delta t}}\exp\Big[-\frac{(L-n-v_{\rm avg}\Delta t)^2}{4D\Delta t}\Big]. \label{Gauss}
\end{equation}
However, in the intermediate time regime, the conditional propagator decays much more slowly than the Gaussian form. Numerical results are shown in Fig.\ref{fig:PLla}, where the free energy continued to be more and more downhill steadily under the same ramping rate.

Similarly, we can calculate the conditional probability $P(l_a,t|L)$ that $l_a$ driven jumps occur when it is known that the avalanche reaches a length $L$. From the full counting statistics, we have that 
\begin{equation}
\begin{aligned}
    P(l_a,t|L)&\approx 
\mathcal Z_L^{-1}\frac{[QR_0(1+e^{-\beta\phi})t]^{l_a}}{l_a!}
e^{-\frac{1}{2}\beta\phi(L-l_a)}\\
&\times
I_{|L-l_a|}\!\Bigl(2\Delta t R_0(1-Q)e^{-\frac{1}{2}\beta\phi}\Bigr),\label{Bessel1}
\end{aligned}
\end{equation}
where the normalization constant $\mathcal Z_L=\sum_{m\ge0}\frac{[QR_0(1+e^{-\beta\phi})t]^{m}}{m!}
e^{-\frac{1}{2}\beta\phi(L-m)}
I_{|L-m|}(2\Delta t R_0(1-Q)e^{-\frac{1}{2}\beta\phi})$. We focus on the tail decay behavior of this conditional probability. By using Stirling's formula and the saddle-point approximation, we find that in the intermediate time scale, rather than having a Gaussian tail, the count probability approximately exhibits an exponential decay, with the slope of the exponent $P(l_a,t|L)\sim e^{\Psi(l_a)}$ being
\begin{equation}
\begin{aligned}
    \frac{d\Psi(l_a)}{d l_a}\Big|_{l_a>L}&=-\ln\Bigl(\frac{L}{2R_0\cosh (\frac{1}{2}\beta\phi)Qt}\Bigr)\\
    &-\Bigl[\ln(1+\frac{\Delta}{L})+\text{arsinh}\big(\frac{\Delta e^{\frac12\beta \phi}}{2tR_0(1-Q)}\big) \Bigr]
\end{aligned}
\end{equation}
with $\Delta =l_a-L$ when $l_a>L$. We observe that the second term changes very slowly when $L$ and $t$ are large. Therefore, the first term provides the base of the exponential tail. 

\begin{figure}[htbp]
  \centering
  \includegraphics[width=0.8\columnwidth]{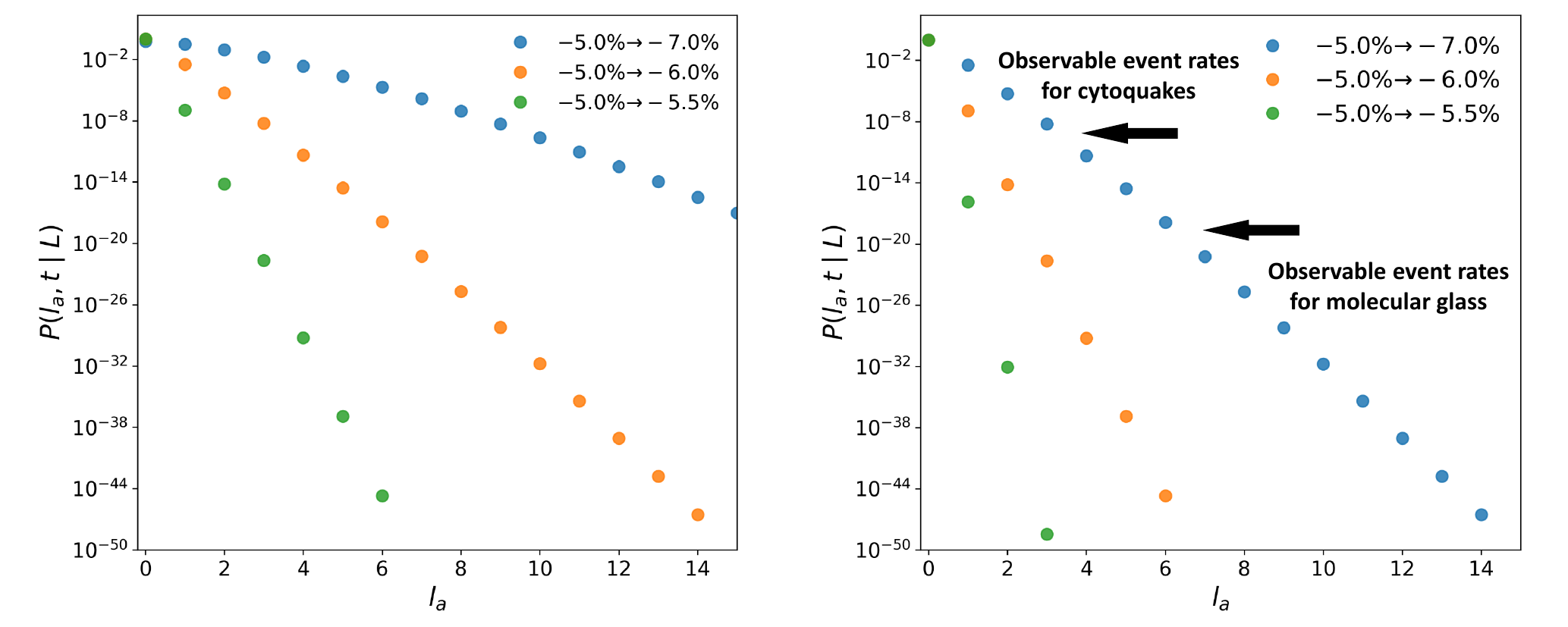}
  \caption{The conditional distribution of avalanches $P(l_a|L,t)$ of size being $L=50$. The system is prepared in the downhill flow situation with a frozen configurational state having $s_c^{\rm frozen}=1.1865k_B$ with the critical string entropy $s_c^{\rm string}=1.13k_B$. At the initial moment, the system was $-5\%$ away from instability. Under the stress ramping condition, the free energy continued to be more and more downhill steadily. Ultimately at time $t$, the system's stability deviation was $-5.5\%$ (green), $-6\%$ (orange), and $-7\%$ (blue). Other parameters are $\Delta C_p=1k_b$, $\eta=0$ and $R_0=0.01s^{-1}$. Left panel: fast ramping $t=100$; Right panel: slow ramping $t=200$. The arrows indicate the ranges, which quantify how microscopic initiation statistics are amplified to experimentally observable avalanche rates in both the cytoquakes and the molecular glass seismology.}
  \label{fig:E}
\end{figure}

\begin{figure}[htbp]
  \centering
  \includegraphics[width=0.8\columnwidth]{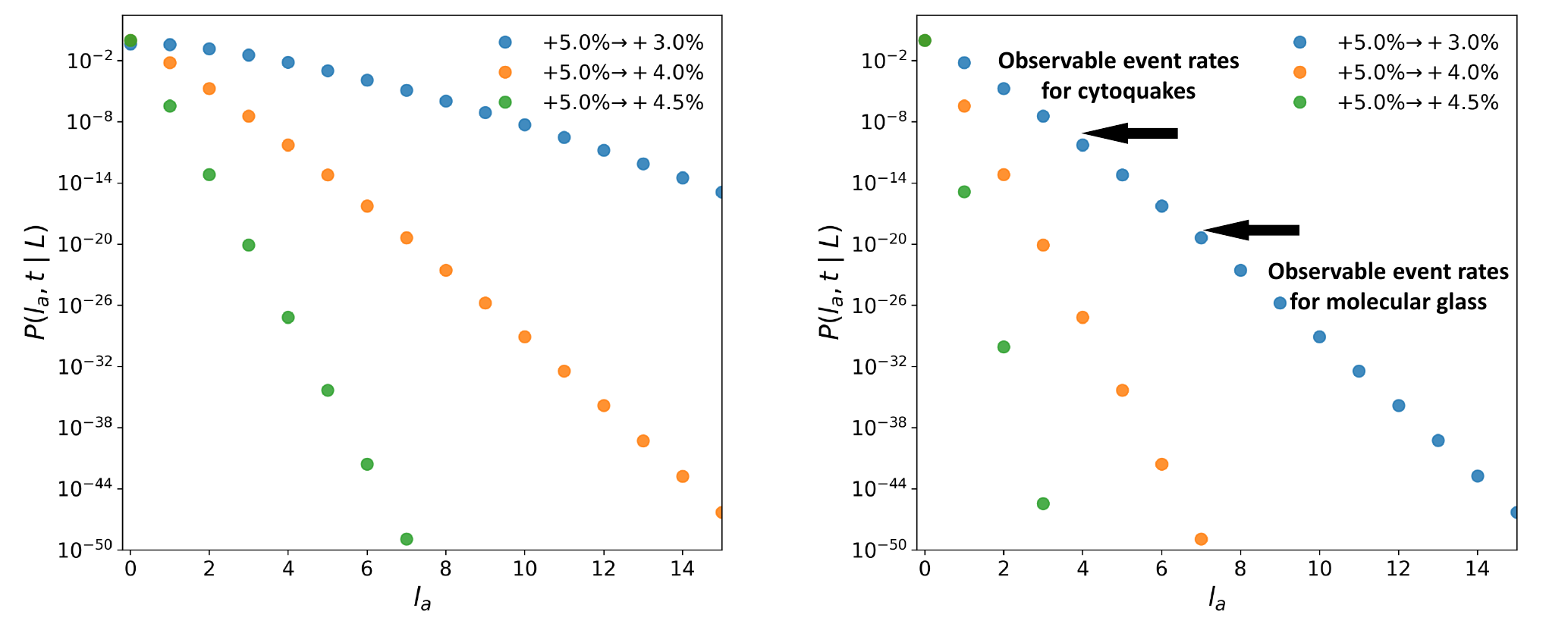}
  \caption{The conditional distribution of avalanches $P(l_a|L,t)$ of size being $L=50$. The system is prepared in the uphill flow situation with a frozen configurational state having $s_c^{\rm frozen}=1.0735k_B$ with the critical string entropy $s_c^{\rm string}=1.13k_B$. At the initial moment, the system was $5\%$ away from instability. Under the stress ramping condition, the free energy becomes more and more downhill steadily. Ultimately at time $t$, the system's stability deviation was $4.5\%$ (green), $4\%$ (orange), and $3\%$ (blue). Other parameters are $\Delta C_p=1k_b$, $\eta=0$ and $R_0=0.01s^{-1}$. Left panel: fast ramping $t=100$; Right panel: slow ramping $t=200$. The arrows indicate the ranges, which quantify how microscopic initiation statistics are amplified to experimentally observable avalanche rates in both the cytoquakes and the molecular glass seismology.}
  \label{fig:F}
\end{figure}

We show the numerical results of  the conditional distribution at intermediate times for both the downhill flow situation [Fig. \ref{fig:E}] and the uphill flow situation [Fig. \ref{fig:F}]. For the downhill flow situation, the system initially starts $-5\%$ away from the instability point and, under stress ramping at different rates, is driven progressively farther from the instability. In contrast, for the uphill flow situation, the system also begins $5\%$ away from instability but is driven closer to the instability point as the stress is ramped. Our simulations show that the probability of avalanche initiation increases with both the degree of downhill driving and the ramping rate, indicating that faster ramping and stronger departure from instability enhance the likelihood of triggering rearrangement events. The numerical results also confirm the exponential tail decay at intermediate times. 

\begin{comment}
    (can be moved to Method/SI Section) Additionally, the full counting statistics at the long-time limit can be obtained uniformly by solving for the dominant eigenvalue. For large $t$, $\ln G(k,z,t)=s_*(k,z)t$ changes linearly with the scalar pole solution $s_*(k,z)$, satisfying $s_*=\lambda(k,s_*;z)$, which is referred to as the scaled cumulant generating function. Inserting this into Eq.\ref{PN|n} and evaluating the Laurent integral by the saddle-point method, one finds that (see \textcolor{blue}{SI} for details),
    $$P(L,t|l_a)\asymp\exp\!\bigl[-t\bigl(I(v,j)-I_0(j)\bigr)\bigr]$$
with
$$j={l_a}/{t},\quad v={L}/{t}$$
the average current of total moves and purely avalanche moves. Here, $I(v,j)=\sup_{k,z}\bigl[ kv+(\ln z)j-s_{*}(k,z)\bigr]$ and $I_0(j)=\inf_v I(v,j)$ are the joint and marginal rate functions, respectively. Such a framework, by combining both the full counting statistics and the large deviation principle, provides a general tool to analyze the avalanche statistics.
\end{comment}

\begin{comment}
    Under shear, a rate-dependent mean-field mobility $\bar{\mu}\!=\!f(\sigma,\partial_t\sigma)\!\approx\! f(\sigma)\!+\!\lambda(\sigma)\partial_t\sigma$ renormalizes the diffusion equation to $[1-\lambda(\sigma)]\partial_t\sigma\!=\!\mu[\nabla^2\sigma+\dots]$. 
\end{comment}

The conditional probability $P(l_a| L,t)$ for a single initiated avalanche (the size $V_{\rm init}\sim L_{\rm init}^3$) can be converted into the probability of observing at least one event within a finite scanning volume $V_{\rm scan}\sim L_{\rm scan}^3$ by introducing the geometric amplification factor $\eta_V \equiv V_{\rm scan}/V_{\rm init}$. Assuming statistically independent initiation volumes and a small local probability, the observed probability is $P_{\rm obs}=1-(1-P_{\rm local})^{\eta_V}\approx \eta_V P_{\rm local}$ with $P_{\rm local}=P(l_a| L,t)$, so multiplying by $\eta_V$ corresponds to the rare-event approximation. In three dimensions, $\eta_V\sim (L_{\rm scan}/L_{\rm init})^3$. For cytoquakes, taking $L_{\rm scan}\sim 1\mu{\rm m} $ and $ L_{\rm init}\sim 1{-}10{\rm nm}$ yields $\eta_V^{\rm cyto}\sim 10^6{-}10^9$. The value $10^9$ corresponds to a system resembling the dendritic spine simulated by Liman et al. \cite{liman2020role} and the study of Floyd et al. \cite{floyd2021understanding}. For soft glassy materials and metallic glasses \cite{cipelletti2000universal,bouchaud2001anomalous,ruta2012atomic}, avalanches appear at random in space and time owing to the long-range elasto-plastic interactions, where a $1{\rm cm}^3$ sample with $L_{\rm init}\sim 3{-}5{\rm nm}$ gives $\eta_V^{\rm glass}\sim 10^{19}{-}10^{21}$. We see that extremely low rates of avalanche initiation at a single slip are indeed observable on the laboratory timescale as indicated in both Figs. \ref{fig:E} and \ref{fig:F}.

\section{Discussion}
The present RFOT-CTRW framework has been formulated under an effectively annealed disorder assumption, where each event samples fresh fluctuations even if the avalanches were to move backward. Sometimes, uphill moves near and below the crossover temperature proceed via rare multi‐step transitions with a lucky sequence of driving force fluctuations \cite{stevenson2010universal}. These multi-step processes can be incorporated into our description via effective coarse-graining, which can interpolate between annealed disorder and quenched disorder \cite{stevenson2010universal}, thereby extending the validity of the present approach across the crossover.

In summary, in this paper, we have presented a unified theory of thermal avalanches in driven glasses by embedding RFOT‐derived string rearrangements within a continuous‐time random‐walk framework.  Our approach naturally yields non‐Poisson waiting‐time distributions that capture both aging under thermal activation and cutoff‐truncated dynamics under external drive, leading to closed‐form effective‐temperature expressions for shear and shaking protocols.  Through large‐deviation and full‐counting‐statistics analyses, we have characterized both long‐time Gaussian limits and rich intermediate‐time, aging‐dominated behavior. This RFOT–CTRW synthesis opens a path to quantitatively link molecular‐scale free‐energy landscapes with experimentally measurable glassy dynamics under diverse loading conditions. 

\bibliographystyle{apsrev4-2}
\bibliography{bibfile}

\appendix

\end{document}